\documentclass[aps,preprint,singlecolumn,nopacs,superscriptaddress,groupedaddress]{revtex4}

\usepackage{amsmath}
\usepackage[dvips]{graphicx}
\usepackage{epsfig}
\usepackage{tabularx}
\usepackage{color}
\usepackage[pdfpagemode=UseNone,colorlinks=true,linkcolor=blue,citecolor=blue]{hyperref}

\usepackage{soul}
\usepackage{gensymb}

\newcommand{\green}[1]{{\color[rgb]{0,0.7,0}#1}}

\newcommand{\di}{i}

\begin{document}

\title[]{Strong coupling and dark modes in the motion of a pair of levitated nanoparticles }

\author{A. Pontin}
\email{antonio.pontin@cnr.it}
\affiliation{CNR-INO, largo Enrico Fermi 6, I-50125 Firenze, Italy}

\author{Q. Deplano}%
\affiliation{Dipartimento di Fisica e Astronomia, Università degli Studi di Firenze, via Sansone 1, I-50019 Sesto Fiorentino, Italy}%
\affiliation{INFN, Sezione di Firenze, via Sansone 1, I-50019 Sesto Fiorentino, Italy}

\author{A. Ranfagni}%
\affiliation{Dipartimento di Fisica e Astronomia, Università degli Studi di Firenze, via Sansone 1, I-50019 Sesto Fiorentino, Italy}%

\author{F. Marino}%
\affiliation{CNR-INO, largo Enrico Fermi 6, I-50125 Firenze, Italy}
\affiliation{INFN, Sezione di Firenze, via Sansone 1, I-50019 Sesto Fiorentino, Italy}

\author{F. Marin}
\email{marin@fi.infn.it}
\affiliation{CNR-INO, largo Enrico Fermi 6, I-50125 Firenze, Italy}
\affiliation{Dipartimento di Fisica e Astronomia, Università degli Studi di Firenze, via Sansone 1, I-50019 Sesto Fiorentino, Italy}%
\affiliation{INFN, Sezione di Firenze, via Sansone 1, I-50019 Sesto Fiorentino, Italy}
\affiliation{European Laboratory for Non-Linear Spectroscopy (LENS), Via Carrara 1, I-50019 Sesto Fiorentino, Italy}

\begin{abstract}
We experimentally investigate a system composed of two levitating nanospheres whose motions are indirectly coupled via coherent scattering in a single optical cavity mode. The nanospheres are loaded into a double longitudinal tweezer created with two lasers at different wavelengths, where chromatic aberration leads to the
formation of two separate trapping sites. We achieve strong coupling between each pair of modes in the transverse plane of the tweezer, as demonstrated by the avoided crossings observed when tuning the eigenfrequencies of the motion of one nanosphere by varying its optical potential depth. Remarkably, we show the emergence of dark modes in the overall coupled motion. The dynamics can be described in terms of spin-1/2 matrices, and the observed features are ubiquitous in a variety of classical and quantum systems. As such, our experiment will allow us to explore the classical analog of typically quantum dynamics, and in further developments to investigate the transition to the quantum domain by lowering the decoherence rate and creating stationary entanglement, as well as implementing non-stationary protocols.
\end{abstract}

\maketitle
Levitated optomechanical systems have shown a fast development over the last decade, and they are a promising platform for the exploration of quantum mechanics in the mesoscopic mass range~\cite{Millen2020Optomechanics,Bassi2013Models}.  Nowadays, the motional ground state for the center of mass of an optically levitated nanoparticle can be routinely prepared, both in free space by feedback ~\cite{Magrini2021,novotny2021_GS,Kamba2022Optical} and through cavity-assisted cooling~\cite{Delic2020science,Ranfagni2021Two-dimensional,Piotrowski2023Simultaneous,Deplano2024Highpurity}. Levitated objects are also seen as attractive sensors both for fundamental science topics, such as dark matter searches~\cite{Monteiro2020Search,Moore2021Searching, Kilian2024Darkmatter} or gravitational wave detection~\cite{Arvanitaki2013Detecting,Aggarwal2022Searching}, and for technological applications~\cite{Monteiro2017Optical,Rademacher2019Quantum,Ahrens2024Levitated}. An additional feature of levitation, compared to clamped optomechanical systems, is the possibility of accessing and manipulating the rotational degrees of freedom, which offer unique opportunities~\cite{Muddassar2018Precession,tongcangli_5D,novotny_libration_fbk2021,Pontin2023Simultaneous,Zielinska2023Controlling}. 

In recent years, an emerging topic within the field has been the study of multiparticle interactions.  Indeed, coupling between two particles has been explored on different platforms including electrodynamic and gravito-optical traps, and optical tweezers. These experiments have demonstrated sympathetic cooling~\cite{Arita2022Alloptical,Lika2023Colddamping, penny2023Sympathetic, Bykov2023_3Dsympathetic} and squeezing~\cite{penny2023Sympathetic}, cold damping~\cite{Vijayan2022Scalable}, optical binding yielding anti-reciprocal coupling~\cite{Rieser2022Tunable} and strong Coulomb coupling~\cite{Deplano2024Coulomb}. This body of work represents the first building block toward more enticing experiments  such as the generation and observation of entanglement between mesoscopic objects~\cite{Chauhan2020Stationary,Brando2021Coherent,Winkler2024Steady-state}, and the exploration of the quantum nature of gravity~\cite{Bose2017spin}.

A recent work ~\cite{Vijayan2024Cavity-mediated} showed strong coupling between pairs of collinear oscillation modes of two levitated nanoparticles. The long range interaction was mediated by an optical cavity in the Doppler regime, exploiting coherent scattering (CS)~\cite{Vuletic2000Laser,novotny_CS_2019,Delic_CS_2019,toros2020Quantum,toros2021Coherent}.
Here, we introduce a different approach, where the two particles are loaded in a bichromatic longitudinal tweezer~\cite{Deplano2024Coulomb}  and the CS interaction operates in the resolved sideband regime. 
This method allows to vary the optical potential independently of the CS coupling term. Furthermore, strong coupling between all modes in the tweezers transverse plane is achieved, as demonstrated by the observed avoided crossings that, for each mode pair, display frequency splittings of $\approx 1$ kHz, significantly exceeding the modes linewidths.

More importantly, we show the emergence of a dark mode in the upper branch of the crossing. 
When considering the motion in a tweezer polarization plane, say $x$ and $y$ modes, it has been previously noted that a single particle interacting by CS with a cavity mode can be described in terms of spin 1 matrices~\cite{toros2021Coherent}, which are usually associated with purely quantum systems. For sufficiently strong coupling, this leads to a three-mode avoided crossing at the tripartite resonance with a dark mode at the center of the crossing~\cite{Ranfagni2021Vectorial}. The same description can be used when considering two modes belonging to different particles.

In the present experiment, we are exploring the weak coupling limit for the interaction of each individual particle with the cavity. This allows to trace out the cavity and describe the dynamics as that of two strongly coupled motional modes \cite{Vijayan2024Cavity-mediated}. Thus, the dynamics can be described in terms of spin 1/2 matrices with an obvious parallel to a quantum two-level system. Indeed, the same features presented here in the classical regime can be found in disparate quantum systems. A first example is a pair of single molecules directly coupled by a dipole-dipole interaction~\cite{Trebbia2022Tailoring}, in which case the bright/dark mode is typically discussed in terms of sub/superradiance. A second example is provided by two transmons qubits coupled by a superconducting cavity, where the bright/dark mode represents the antisymmetric/symmetric superposition of the two qubit states~\cite{Majer2007Couplingsuperconducting}. Other notable examples can be found in Refs.~\cite{Evans2018Photon-mediated,Filipp2011Multimode,Leseleuc2017Optical}.

\begin{figure}
\includegraphics[width=8.6cm]{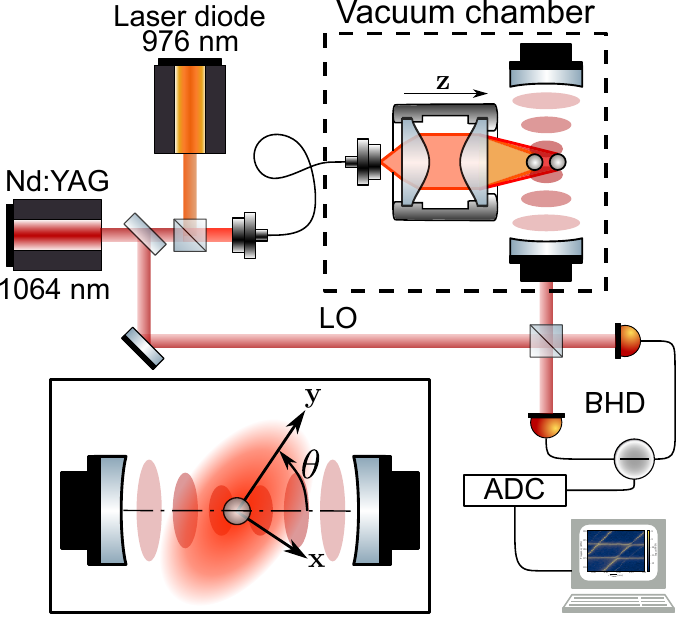}
\caption{Overview of the experiment. A bi-chromatic optical tweezer allows to confine nanoparticles in two trapping sites separated by $r_{12}=9\pm1\,\mu$m. The two particles are placed at the center of a high finesse optical cavity implementing a coherent scattering approach. The particles motion is measured through a balanced heterodyne detection (BHD). The tweezer polarization is set at an angle $\theta\approx45$°, giving similar optomechanical couplings for the $x$ and $y$ directions. LO: local oscillator, ADC: analog-to-digital converter.  
}\label{fig1}
\end{figure}

\emph{Experimental setup} - We implement a fiber-based approach to obtain our optical tweezer. Light from a polarization maintaining (PM) fiber is first collimated by an aspheric lens (focal length $f=18.4$\,mm) and subsequently refocused by a second asphere ($f=3.1$\,mm). This generates a sub-$\mu$m waist. Silica particles are initially loaded in an auxiliary tweezer housed in a loading chamber, and then transferred to the main tweezer~\cite{Calamai2021Transfer}. The main aspheric doublet is mounted on a 3 axes translational stage which allows nanometric positioning at the center of an optical cavity.  

We inject light into the PM fiber from two sources, a Nd:YAG at $1064$\,nm  and a diode laser at $976$\,nm, which are overlapped at the fiber input with orthogonal polarizations. Chromatic aberration of the doublet guaranties that the foci occur at different positions along the tweezer field propagation direction ($z$ axis). For our wavelengths, the separation between the two spots is estimated to be $r_{12}=9\pm1\,\mu$m. 

The optical cavity has a quasi-concentric configuration with identical concave mirrors, a free spectral range FSR$=3.07$\,GHz and a waist of $31\,\mu$m. The cavity full linewidth is $\kappa/2\pi=57$\,kHz at $1064$\,nm while the cavity mirrors are effectively transparent at $976$\,nm. A second Nd:YAG locked to the optical cavity provides a reference to offset phase-lock the $1064$\,nm tweezer light. The frequency offset is set at a frequency of FSR$+\Delta/2\pi$ allowing precise control of the tweezer field detuning $\Delta$ with respect to the cavity resonance.

\emph{Model} - Both particles couple to the cavity mode through a CS interaction ~\cite{Vuletic2000Laser,novotny_CS_2019,Delic_CS_2019,toros2020Quantum,toros2021Coherent}. Light directly scattered by the particles is injected into the cavity mode giving rise to a coupling between the intracavity field and the particles motion.  Contrary to the typical scenario~\cite{Vijayan2024Cavity-mediated}, the light generating the second trap, at $976$\,nm, does not participate in the coherent scattering interaction. However, the particle in this trap still scatters $1064$\,nm light into the cavity albeit with reduced power. We can write the electric field of the $1064$\,nm light at particle 2 as $E_2=\zeta\,E_1\,e^{i\varphi}$. Here, $E_1$ is the field at the $1064$ focus while $\zeta$ and $\varphi$ account for the different amplitude and phase at the $976$ focus.  These two parameters fully characterize the relative strength of the CS interaction for the second particle. Based on a Gaussian approximation of the field propagation we can estimate $\zeta=0.28\pm0.03$. Notice that $\zeta<1$ due to the geometry of the system.

The experimental results presented here explore only the weak coupling limit of the interaction between motion and cavity field. In this regime, modeling of the experiment simplifies considerably, since the cavity can be traced out. It can be shown \cite{Vijayan2024Cavity-mediated,supp} that the cavity mediated coupling between the two particles assumes the following form 

\begin{equation}\label{eq_2} 
   G_{\alpha\beta}=\frac{g_\alpha g^*_{\beta}}{\Delta - \omega_\beta - \mathrm{i} \kappa/2} + \frac{g^*_\alpha g_{\beta}}{\Delta + \omega_\beta +\mathrm{i} \kappa/2}
\end{equation}

\noindent where $g_i$ is the coupling rate between motional mode \emph{i} and cavity field, and $\omega_i$ is the modal eigenfrequency. Eq.~(\ref{eq_2}) allows to focus directly on the motional dynamics of the two particles. In general, $G_{\alpha\beta}\neq G_{\beta\alpha}$ so that the interaction can be non-reciprocal. Maximal non-reciprocity occurs at a phase $\varphi=\pi/2$, whereas a purely conservative interaction is recovered for $\varphi=0$.   

\begin{figure}
\includegraphics[width=8.6cm]{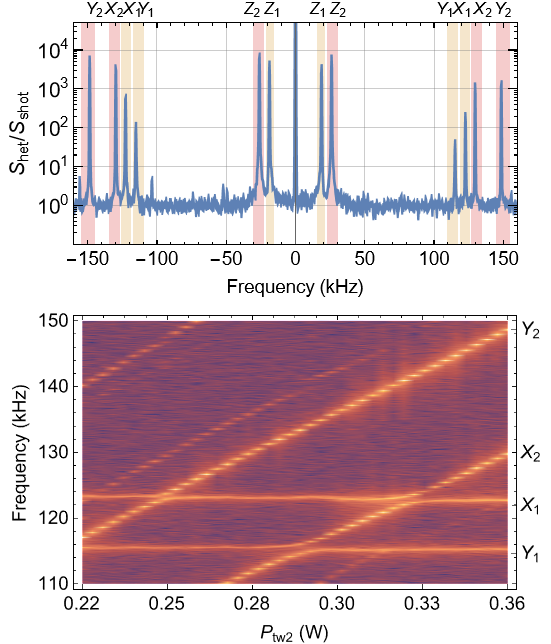}
\caption{Experimental spectra. a) Heterodyne spectrum of the cavity output normalized to the shot noise. The spectral peaks of the two particles degrees of freedom $(z_1,x_1,y_1,z_2,x_2,y_2)$ are identified with yellow and red shaded regions for particle 1 and 2, respectively. b) Spectrogram of the heterodyne power spectral density (PSD) as a function of the trapping power $P_{\text{tw2}}$ of the $976$\,nm field. The power is varied in discrete steps. Three avoided crossings are clearly visible at $P_{\text{tw2}}=(0.32,0.29,0.24)$\,W. This demonstrates an effective direct coupling between mechanical modes belonging to different particles.
}\label{fig2}
\end{figure}

\emph{Experiment} - We trap two silica particles of nominal diameter $125$\,nm. The power of the $1064$\,nm field is kept constant throughout the experiment at $P_{\mathrm{tw1}}=250$\,mW. This gives trap frequencies of $(\omega_{z1},\omega_{y1},\omega_{x1})/2\pi=(19,117,124)$\,kHz for the first particle. The power of the $976$\,nm light can be varied  from $P_{\mathrm{tw2}}=360$\,mW down to $220$\,mW. The lower bound is set to avoid particle loss. At the highest power, we have $(\omega_{z2},\omega_{y2},\omega_{x2})/2\pi=(28,148,129)$\,kHz, and the frequency difference ensures that  the two particles can be considered as independent. The $1064$\,nm field polarization is linear and set at $\theta\simeq45\degree$, where $\theta$ is the angle between the $y$ direction and the cavity axis, so that we expect $g_{x_i}\simeq g_{y_i}$. We show in Fig.~\ref{fig2}\,a a heterodyne spectrum taken at a pressure of $5\times10^{-6}$\,mbar with a red detuning set at $\Delta/2\pi=-425$\,kHz. All spectral peaks of the two particles degrees of freedom are clearly recognizable. The two particles are charged with measured charges of $ |q_1|= (89\pm10)\,q_e$ and $|q_2|= (71\pm10)\,q_e$ where $q_e$ is the elementary charge~\cite{Deplano2024Coulomb}. 

The two particles are positioned close to the cavity axis, and the position along the cavity standing wave is set by minimizing the total light scattered into the cavity mode. This does not necessarily occur at a cavity node as is typical in single particle experiments~\cite{supp}.

By controlling the power of the $976$\,nm field we can tune its trap frequencies to be resonant to those of the $1064$\,nm. Since the foci parameters are sufficiently different, the resonance condition is achieved only for a single pair of modes at a time. We show in Fig.~\ref{fig2}\,b) a spectrogram obtained by scanning $P_{tw2}$, displaying  the frequency region around the resonances of the modes in the tweezers transverse plane (i.e., $x_i,y_i$). The power is changed in discrete steps of $3.5$\,mW. Three avoided crossings are clearly observable, and a fourth remains just outside the scanning range. Fitting with Lorentzian shapes the four spectral peaks at each step, we reconstruct the behavior of the four eigenfrequencies of the system.

\begin{figure}[ht]
\includegraphics[width=8.6cm]{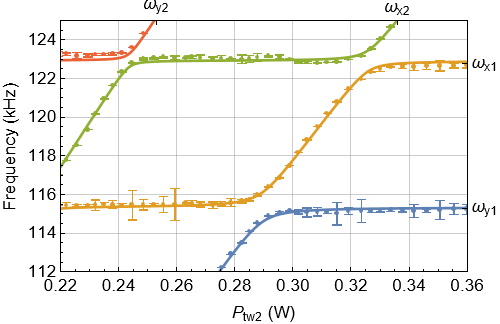}
\caption{Measurement of the system eigenfrequencies for the motion in the tweezers transverse plane. The experimental data (markers with error bar) are shown along with the modeled eigenfrequencies (solid lines). Three avoided crossings are visible, allowing to infer effective direct coupling rates of $(g_{x1,x2}, g_{y1,x2}, g_{x1,y2})/2\pi=(0.61, 0.6, 0.51)$\,kHz respectively.
}\label{fig3}
\end{figure}

We model the experiment considering nominal particle and cavity parameters, in line with previous experiments~\cite{Ranfagni2021Two-dimensional}, and fit the measured frequency splittings to extract the cavity drive ratio $\zeta$, relative field phase $\varphi$ and the initial position in the cavity standing wave~\cite{supp}. The resulting eigenfrequencies are shown in Fig.~\ref{fig3} together with the experimental data, demonstrating excellent agreement.

The fitting procedure returns a phase $\varphi=(172\pm 3) \degree$, thus very close to optimal for a purely conservative interaction, and a coupling strength $\zeta=0.35 \pm 0.015$ which is in fairly good agreement with the crude estimate using a Gaussian approximation.  For the particle in the $1064$\,nm trap we find $(g_{x1}, g_{y1})/2\pi=(18.1\pm 0.6, 17.3\pm 0.5)$\,kHz, while for the second particle we have $(|g_{x2}|, |g_{y2}|)/2\pi=(P_{0}/P_{tw2})^{1/4} (7.0\pm 0.3, 6.0\pm 0.2)$\,kHz where $P_{0}=220$\,mW is a reference power.

\emph{Discussion} - The cavity mediated interaction between the two particles can be described as an effective direct coupling $g_{\alpha,\beta}=\mid G_{\alpha,\beta} \mid_{\omega_{\alpha}=\omega_{\beta}}$. The normal mode splitting at the avoided crossings is then given by $\delta_{\alpha,\beta}=2 g_{\alpha,\beta}$. For the three visible crossings, occurring at a power of $P_{\mathrm{tw2}}=(326,291,243)$\,mW, we find $(\delta_{x1,x2}, \delta_{y1,x2}, \delta_{x1,y2})/2\pi=(1.22\pm 0.06, 1.20\pm 0.06, 1.02\pm 0.06)$\,kHz respectively. The fourth splitting, occurring at $P_{\mathrm{tw2}}=214$\,mW, can be estimated to be $\delta_{y1,y2}/2\pi=0.78\pm 0.05$\,kHz.

In the case of interacting modes of different linewidths \green{$\gamma_i$}, we can define the strong coupling threshold as $g_{\alpha,\beta} > (\gamma_\alpha +\gamma_\beta)/4$. The largest measured linewidth, dominated by the optomechanical interaction, is  $\gamma_{\text{opt}}/2\pi =109\pm 20$ \,Hz. Taking this as upper-bound ensures that all crossings are well within the strong coupling regime. 

\begin{figure}[t]
\includegraphics[width=8.6cm]{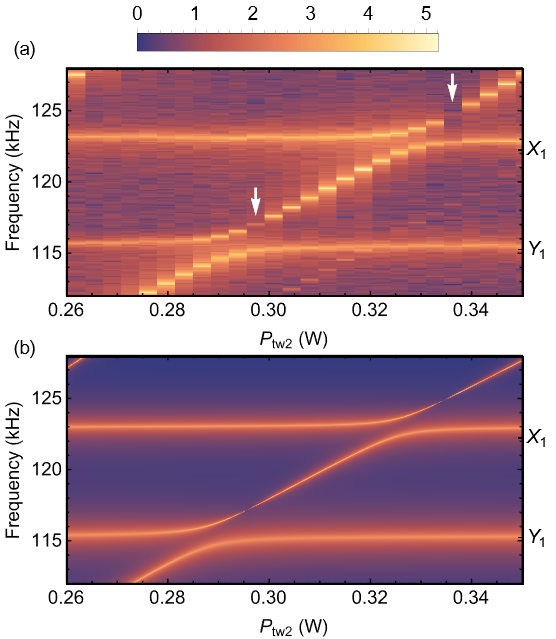}
\caption{Dark mode formation. a) Spectrogram of the Heterodyne PSD as a function of the trapping power $P_{\text{tw2}}$ of the $976$\,nm field around the avoided crossings at $P_{\text{tw2}}=(323,289)$\,mW which shows the formation of dark modes, marked with arrows, on the upper branches. b) analytical heterodyne spectrogram calculated with fitted parameters with the only exception of setting the phase $\varphi=\pi$ to highlight the dark mode.
}\label{fig4}
\end{figure}

To understand the formation of the dark mode at an avoided crossing, let us first consider the simplified scenario of two motional modes of frequencies $\omega_1=\omega_0-\delta \omega$ and $\omega_2=\omega_0+\delta \omega$ with identical coupling rates to the cavity $g=g_1=g_2$, which implies $\varphi=0$. In the large detuning limit, i.e. $|\Delta|\gg \kappa, \omega_j$, the eigenvalues of the dynamical matrix~\cite{supp} at the crossing ($\delta\omega=0$) are given by $(\lambda_+,\lambda_-)=(\omega_0,\,\omega_0+4g^2/\Delta+\di\,4\kappa g^2 \omega_0/\Delta^3)$. It is clear that one mode is down-shifted and broadened (lower branch $\lambda_-$) while the other is completely unaffected by the cavity (upper branch $\lambda_+$) and can then be identified as the dark mode. However, when a gain imbalance is present, the real and imaginary parts of the eigenvalues are affected differently. If we now consider $g_2=\zeta\,g$, the minimal distance between the pair of eigenfrequencies, defining the position of the avoided crossing, occurs at $\delta\omega=(1-\zeta^2)g^2/\Delta$ where $\mathrm{Re}[\lambda_{\pm}]=\omega_0+(\zeta\mp1)^2 g^2/\Delta$, but it is still at $\delta\omega=0$ that the imaginary part of $\lambda_+$ vanishes, and $\mathrm{Im}[(\lambda_+,\lambda_-)]=(0,\,(1+\zeta^2)\,2\kappa g^2 \omega_0/\Delta^3)$. In other words, the dark mode always forms at the crossing of the bare frequencies, not at the position of minimal distance between optically-shifted effective frequencies.

In fact, this is the scenario that can be observed in the experimental spectrogram of Fig.~\ref{fig4}~a). For both avoided crossings shown, the amplitude of the upper branch significantly dims at a power corresponding to a $\delta\omega$ higher than that of the crossing, as expected. The amplitude does not completely vanish for three reasons: \emph{i}) the dark mode only appears for $\varphi=0$, away from this value one should rather consider a "pseudo"-dark mode, with a contrast quickly vanishing as the phase moves away from the optimal value. \emph{ii}) The region over which the dark mode manifests itself is narrow compared to the resolution of the power scan. \emph{iii}) Due to the Coulomb force between the charged particles, changing the trap potential results in a shift of the equilibrium position, i.e., the particles separation, of the order of few $100$\,nm. This introduces a power dependence on the relative phase of the fields $\varphi$, though with negligible effect on $\zeta$. For comparison, we show in Fig.~\ref{fig4}\,b) the analytical heterodyne spectrogram. This is calculated with the fitted parameters except for the phase $\varphi$ and the effect of the Coulomb shift, which have been both set to zero to show the maximum contrast of the dark mode. 

As mentioned in the previous paragraph, the Coulomb force is not negligible, due to its effect on the position-dependent field phase. Coulomb coupling affects the dynamics of the mode pairs $(x_1,x_2)$ and $(y_1,y_2)$ with a coupling rate estimated to be $g_C/2\pi\simeq120$\,Hz, and it thus introduces a minor contribution to the corresponding avoided crossings. To model the eigenfrequencies shown in Fig.~\ref{fig3}, both the direct effective couplings and the Coulomb force are fully included in the model. More details can be found in~\cite{supp}.  On the other hand, at the separation $r_{12}=9\,\mu$m optical binding can be neglected.

In conclusion, we exploited a novel scheme to confine two nanoparticles in separated trapping sites of a double optical tweezer, and place them at the center of a high finesse optical cavity. This allowed us to exploit a coherent scattering interaction to generate an effective direct coupling between interparticle modes, mediated by the cavity. We have directly shown the appearance of three avoided crossings between pairs of motional modes in the tweezer polarization plane with splittings of $\approx1$\,kHz. A fourth remains just outside our scanning range, but is predicted to be of the same order. All crossings are well within the strong coupling regime. We have also shown that the effective direct coupling leads to the formation of a dark mode in the upper branch of the avoided crossing spectrogram, which occurs when the bare frequencies are degenerate.

These features are quite common across a variety of physical systems, including those operating deeply in the quantum regime~\cite{Trebbia2022Tailoring,Majer2007Couplingsuperconducting}. Indeed, an experiment in the classical regime can represent the classical analog of quantum dynamical phenomena including, for example, Rabi oscillations and Ramsey fringes~\cite{Ivakhnenko2018Simulating,Frimmer2014TheclassicalBloch}, which could be explored with our system. Of course, an even more interesting scenario would be achieved by preparing the two particles in their mechanical ground state to pursue the generation of stationary Gaussian entanglement \cite{Chauhan2020Stationary}, as well as non-stationary quantum protocols \cite{Rudolph2020,Brando2021Coherent}. To achieve this, some upgrades for the current setup will be necessary. However, these are fully achievable with the technology currently available. For example, a widely tunable light source for the second tweezer will allow control of the interaction phase $\varphi$ and particles separation $r_{12}$, which directly affects the interaction strength $\zeta$. It is also possible to envision a regime where both cavity-mediated interaction and optical binding are relevant. Lastly, a doubly resonant cavity at both $1064$\,nm and $976$\,nm, would represent a novel approach within the CS framework with a much larger parameter space available to tailor the interaction.

We acknowledge financial support from PNRR MUR Project No. PE0000023-NQSTI and by the
European Commission-EU under the Infrastructure I-PHOQS “Integrated Infrastructure Initiative
in Photonic and Quantum Sciences ” [IR0000016, ID D2B8D520, CUP D2B8D520].

\bibliographystyle{nicebib}
\bibliography{main_bib}

\end{document}


\title[]{Supplementary Material for: Strong coupling and dark modes in the motion of a pair of levitated nanoparticles}

\author{A. Pontin}
\email{antonio.pontin@cnr.it}
\affiliation{CNR-INO, largo Enrico Fermi 6, I-50125 Firenze, Italy}

\author{Q. Deplano}%
\affiliation{Dipartimento di Fisica e Astronomia, Università degli Studi di Firenze, via Sansone 1, I-50019 Sesto Fiorentino, Italy}%
\affiliation{INFN, Sezione di Firenze, via Sansone 1, I-50019 Sesto Fiorentino, Italy}

\author{A. Ranfagni}%
\affiliation{Dipartimento di Fisica e Astronomia, Università degli Studi di Firenze, via Sansone 1, I-50019 Sesto Fiorentino, Italy}%

\author{F. Marino}%
\affiliation{CNR-INO, largo Enrico Fermi 6, I-50125 Firenze, Italy}
\affiliation{INFN, Sezione di Firenze, via Sansone 1, I-50019 Sesto Fiorentino, Italy}

\author{F. Marin}
\email{marin@fi.infn.it}
\affiliation{CNR-INO, largo Enrico Fermi 6, I-50125 Firenze, Italy}
\affiliation{Dipartimento di Fisica e Astronomia, Università degli Studi di Firenze, via Sansone 1, I-50019 Sesto Fiorentino, Italy}%
\affiliation{INFN, Sezione di Firenze, via Sansone 1, I-50019 Sesto Fiorentino, Italy}
\affiliation{European Laboratory for Non-Linear Spectroscopy (LENS), Via Carrara 1, I-50019 Sesto Fiorentino, Italy}


\maketitle

\tableofcontents

\section{Model}
As in standard description of coherent scattering (CS)~\cite{Vuletic2000Laser,novotny_CS_2019,Delic_CS_2019,toros2020Quantum,toros2021Coherent} we can write the total optical potential as $H_{\text{opt}}=- \mid \sqrt{\alpha_1} [\mathbf{E}_{1}(\mathbf{r}_1)+\mathbf{E}_{\text{cav}}(\mathbf{r}_1)]+\sqrt{\alpha_2}[\mathbf{E}_{2}(\mathbf{r}_2)+\mathbf{E}_{\text{cav}}(\mathbf{r}_2)]\mid^2/4$, where $\mathbf{r}_i$ and $\alpha_i$ are the two particles positions and their polarizability, $\mathbf{E}_i$ are the electric fields near the foci of the two trapping sites and $\mathbf{E}_\text{cav}$ is the intracavity field. As described in the main text, the second tweezer light (at $976$\,nm) is not resonant with the cavity so it does not participate in the CS interaction. The second particle interacts with the cavity mode through the scattered $1064$\,nm light from the first tweezer, which we write as $E_1(\mathbf{r}_2)=\zeta\,E_1(0)\,e^{i\varphi}$ where $\zeta$ and $\varphi$ account for the different amplitude and phase at the $976$\,nm focus. Without loss of generality, we will assume two identical particles to simplify the notation. We will also neglect the standard dispersive coupling which has been shown to be significantly smaller than the CS couplings.

Before proceeding, it is useful to define the reference frame. We take the $1064$\,nm tweezer propagating along the \textit{z}-axis and with a polarization vector along the \textit{y}-axis. These directions are identified by the versors $\textbf{e}_z$ and $\textbf{e}_y$ respectively. The $976$\,nm tweezer is polarized along $\textbf{e}_x$. As a starting point for our definition, we consider a standard configuration where the cavity axis lies along $\textbf{e}_y$ and the cavity mode is polarized along $\textbf{e}_x$. In our setup, both tweezers polarizations are then rotated by an angle $\theta$ (see Fig. \ref{fig3sup}). However, it is convenient to keep the reference frame fixed on the tweezer and consider a rotated cavity frame. After rotation, the cavity mode is polarized along $\mathbf{e}_1$ and its axis lies along $\textbf{e}_2$. We also need to consider a deviation from orthogonality between the tweezer propagation direction and the cavity axis by a small angle $\theta_z$ to reflect an experimental imperfection. With this we can define $\mathbf{e}_1=\mathcal{R}(-\theta,\mathbf{e}_z)\mathcal{R}(-\theta_z,\mathbf{e}_x)\mathbf{e}_x$ and $\mathbf{e}_2=\mathcal{R}(-\theta,\mathbf{e}_z)\mathcal{R}(-\theta_z,\mathbf{e}_x)\mathbf{e}_y$ where $\mathcal{R}(\theta,\mathbf{e}_j)$ is a 3D rotation matrix around $\mathbf{e}_j$.

The optical Hamiltonian under consideration can then be written as

\begin{align}\label{eq1s} 
   H_{\text{opt}} & = \frac{\alpha}{4} \biggl[ \mid  \mathbf{E}_{1}(\mathbf{r}_1) \mid^2 + \mid \mathbf{E}_{2}(\mathbf{r}_2) \mid^2 \biggr. \nonumber\\
   &\biggl. + (\hat{a} E_d e^{-\di\mathbf{k}\cdot \mathbf{r}_1} f_{\text{c}} (\mathbf{r}_1) \mathbf{e}_1 \cdot \mathbf{e}_y  + \hat{a} \zeta E_d e^{-\di\mathbf{k}\cdot\mathbf{r}_2+\di\varphi} f_{\text{c}} (\mathbf{r}_2) \mathbf{e}_1 \cdot \mathbf{e}_y + h.c. )   \biggr].
\end{align}
Here, the first two terms are the 3D trapping potentials of the two particles, while the third term represents the coherent scattering interaction~\cite{Vuletic2000Laser,toros2021Coherent}. In Eq.~\ref{eq1s}, $E_d$ is the field driving the cavity from the first tweezer, i.e., $E_d\equiv E_1(0)$, $\mathbf{k}$ is the $1064$\,nm light wavevector, $\hat{a}$ is the cavity mode operator and $f_{\text{c}}(\mathbf{r}_i)$ is the cavity standing wave at the particles position which can be written as

\begin{equation}\label{eq2s}
    f_{\text{c}}(\mathbf{r}_i)=\epsilon_c^2 \cos(k\,\mathbf{e}_2\cdot \mathbf{r}_i+\phi_i)
\end{equation}

\noindent where we have neglected the Gaussian envelope of the cavity mode since $w_c\gg r_{12}$ (with $w_c$ the cavity waist). In Eq.~\ref{eq2s}, $k$ is the $1064$\,nm light wave number, $\phi_i$ identifies the particles position in the standing wave and we have included a normalization factor $\epsilon_c=\sqrt{2\hbar \omega_l/(\epsilon_0 V_c)}$ where $V_c=\pi w_c^2 L_c/4$ is the cavity mode volume, $L_c$ its length and $\omega_l$ is the laser angular frequency. If $\theta_z=0$ then $\phi_1=\phi_2$. In the case of a small, yet non-null $\theta_z$, since $\mathbf{r}_2=(x_2,y_2,z_2-r_{12})$, we can write  $\phi_2=\phi_1+k\,r_{12}\theta_z$. Considering that $r_{12}\simeq9\,\mu$m, even a small $\theta_z$ can have a significant impact on $\phi_2$.

Since the two particles are charged, the Coulomb potential must be considered as well. This is given by

\begin{equation}
    H_C=\frac{q_1 q_2}{4\pi\epsilon_0}\frac{1}{\sqrt{(r_{12}+z_1-z_2)^2+(y_1-y_2)^2+(x_1-x_2)^2}}
\end{equation}

\noindent where $q_i$ is the total charge on particle \emph{i}. We will show in Sec.~\ref{subscoulomb} that the dynamical Coulomb coupling provides only a minor contribution to the avoided crossings, however, one has to take into account the steady state shift along the \emph{z} direction as the power of  the second tweezer is varied to tune the eigenfrequencies of the system. To lowest order the steady state mean positions are given by

\begin{align}\label{eqs_steady}
    \bar{z}_1&= \frac{k_c r_{12}}{2 k_c (1+\omega_{z1}^2/\omega_{z2}^2)+r_{12}^3 m \omega_{z1}^{2}}\\
    \label{eqs_steady1}
    \bar{z}_2&= -\frac{k_c r_{12}}{2 k_c (1+\omega_{z2}^2/\omega_{z1}^2)+r_{12}^3 m \omega_{z2}^{2}}.
\end{align}

\noindent where we have defined $k_c=q_1q_2/(4\pi\epsilon_0)$ to simplify the notation. Although it is notoriously difficult to measure an absolute steady state shift, here only the change  related to the scan of $\omega_{z2}$ is relevant. To take it into account, we define $\Delta z$ to be the variation of the particles separation taking as a reference a given initial condition. The main consequence of $\Delta z$ is a variation of $\varphi$ while sweeping $\omega_{z2}$.

At this point, we can expand to second order the optical Hamiltonian and, moving to dimensionless units by normalizing the displacements to the respective zero point fluctuations ($zpf$), we obtain the following couplings between the cavity mode and the motional degrees of freedom

\begin{align}
    g_{x1}&\simeq -x_{1,zpf}\, g_0 \sin^2{\theta} \sin{\phi_1} \\
    g_{y1}&\simeq -y_{1,zpf}\, g_0 \cos{\theta}\sin{\theta} \sin{\phi_1} \\
    g_{z1}&\simeq -z_{1,zpf}\, g_0 \sin{\theta} (\di\cos{\phi_1}-\theta_z\sin{\phi_1})\\
    g_{x1y1}&\simeq - 2x_{1,zpf}\,y_{1,zpf}\, k\, g_0 \,\alpha_R \cos{\theta}\sin^2{\theta}\\
    g_{x2}&\simeq -x_{2,zpf}\, \zeta \,g_0 \,e^{\di(\varphi-k\,\Delta z)} \sin^2{\theta} \sin(\phi_1+k\,\theta_z r_{12}) \\
    g_{y2}&\simeq -y_{2,zpf}\, \zeta \,g_0 \, e^{\di(\varphi-k\,\Delta z)} \cos{\theta} \sin{\theta}\sin(\phi_1+k\,\theta_z r_{12}) \\
    g_{z2}&\simeq -z_{2,zpf}\, \zeta \,g_0 \, e^{\di(\varphi-k\,\Delta z)} \sin{\theta}[\di\cos(\phi_1+k\theta_z r_{12})-\theta_z\sin(\phi_1+k\theta_z  r_{12})]\\
    g_{x2y2}&\simeq - 2 x_{2,zpf}\,y_{2,zpf} \,\zeta\, k\, g_0 \cos \theta \cos(\phi_1+k\theta_z r_{12})\sin^2\theta\,[\alpha_R\cos(k \Delta z-\varphi)+\alpha_I \sin(k\Delta z-\varphi)]
\end{align}

\noindent where we have defined $g_0=\alpha\epsilon_ckE_d/(4\hbar)$. The steady-state intra-cavity field $\alpha_S=\alpha_R+\di \,\alpha_I$ is given by

\begin{equation}
    \alpha_S=\frac{g_0}{k}\frac{\cos \phi_1+\zeta e^{i (k \Delta z-\varphi)}\cos(\phi_1+k\theta_z r_{12})}{\Delta+ \di \,\kappa}\cos \theta.
\end{equation}

Since the experiment explores only the weak coupling limit, it is mathematically convenient to apply the rotating wave approximation for the mechanical degrees of freedom by introducing their creation and annihilation operators, and to adiabatically eliminate the cavity~\cite{Vijayan2024Cavity-mediated}. In this way one can reduce the dimensionality of the description and obtain an effective direct coupling between the mechanical oscillators. Furthermore, motion along the tweezers propagation direction (\emph{z}$_i$-axis) is largely decoupled from the motion in the transverse planes due to largely different oscillation frequencies. Thus, one can focus on the particles motion in their respective polarization planes. Finally, in our experimental configurations the particles are positioned close to a minimum of the intracavity field, therefore the direct couplings quantified by $g_{x1y1}$ and $g_{x2y2}$ have a negligible role and will be neglected in the following description.

The evolution equations for the cavity field $a$ and the mechanical operators $b_i$ can be written as 
\begin{eqnarray}
\dot{a} &=& \left(i\Delta-\frac{\kappa}{2}\right) a - i \Sigma_\alpha g_\alpha (b_\alpha + b^{\dagger}_\alpha) 
\label{eq_adot}\\
\dot{b}_\alpha &=& \left(-i\omega_\alpha-\frac{\gamma}{2}\right) b_\alpha -i (g_\alpha a^{\dagger}+g_\alpha^* a)
\label{eq_bdot}
\end{eqnarray}
where  the indexes vary though $(x_1,y_1,x_2,y_2)$, $\gamma$ is the mechanical damping rate and we are neglecting for the moment the input noise sources. In Eq. \ref{eq_bdot} the meaningful field terms are those rotating at the mechanical frequencies $\omega_\alpha$, while counter-rotating terms can be neglected. Therefore, from Eq. \ref{eq_adot} and its Hermitian conjugate we write 
\begin{eqnarray}
a &\simeq& \Sigma_\alpha\, \chi_c (\omega_\alpha)\, g_\alpha b_\alpha  
\label{eq_a}\\
a^{\dagger} &\simeq&  \Sigma_\alpha\, \chi_c^* (-\omega_\alpha) \,g_\alpha^* b_\alpha  
\label{eq_adag}
\end{eqnarray}
where $\chi_c (\omega) = 1/(\Delta + \omega +i \kappa/2)$ is the cavity susceptibility. Replacing Eqs. \ref{eq_a}-\ref{eq_adag} in Eq. \ref{eq_bdot} we obtain the cavity mediated coupling rates which, as reported in the main text, are given by
\begin{equation}\label{eq4s}
   G_{\alpha\beta}=\frac{g_\alpha g^*_{\beta}}{\Delta - \omega_\beta - \di\, \kappa/2} + \frac{g^*_\alpha g_{\beta}}{\Delta + \omega_\beta +\di\, \kappa/2}.
\end{equation}

The dynamics can then be described in compact form as

\begin{equation}\label{eqs_dyn}
    \frac{d}{dt}\mathbf{V}=-\di\,\mathbf{D}\mathbf{V}+\mathbf{V}_N
\end{equation}
\noindent where $\mathbf{V}=(b_{x1},b_{y1},b_{x2},b_{y2})$ is the variables vector, $\mathbf{V}_N$ the heating rate vector in which, in our experimental conditions, the only relevant contribution is thermal noise, and the dynamical matrix $\mathbf{D}$ is given by

\begin{equation}\label{eq3s}
    \mathbf{D}_{\alpha\beta}=\bigg[\omega_\alpha-\di\frac{\gamma}{2}\bigg]\delta_{\alpha\beta}+G_{\alpha\beta}.
\end{equation}

\subsection{Eigenvalues near an avoided crossing}\label{sec_eigen}
An eigenvalues analysis of the matrix $\mathbf{D}$ provides all the relevant information regarding the 2 particles dynamics. Even if fully analytical expressions can be calculated, these become cumbersome very quickly, and often a numerical approach is preferable. It is instructive to consider a single avoided crossing between two general modes of frequencies $\omega_1=\omega_0-\delta\omega$ and  $\omega_2=\omega_0+\delta\omega$ that are coupled according to Eq.~\ref{eq4s}. Without loss of generality, we neglect the gas damping $\gamma$.  The eigenvalues are then given by
\begin{equation}
    \lambda_{\pm}=\frac{1}{2}\bigg(2  \omega_0+G_{11}+G_{22}\pm\sqrt{4G_{12}G_{21}+(2\delta\omega+G_{22}-G_{11})^2}\bigg).
\end{equation}
Even in this case, expressions can be rather long so we report some meaningful approximated results. We take a purely conservative interaction with $g_2=\zeta g_1$, and we consider small variations of the frequencies around $\omega_0$, so that in $G_{ij}$ we set $\omega=\omega_1=\omega_2$ and by defining 
\begin{equation}
    G_{00}= g_1^2 \left[ \frac{1}{\Delta-\omega_0+\di\kappa/2}+\frac{1}{\Delta+\omega_0-\di\kappa/2} \right]
\end{equation}
we can write $G_{11}=G_{00}$, $G_{12}=G_{21}=\zeta G_{00}$, and $G_{22}=\zeta^2 G_{00}$.
Moreover, the experimental parameters allow us to expand the eigenvalues assuming a large detuning limit, i.e., $\mid\Delta\mid\gg\kappa,\omega_i$, therefore we are also giving simplified expressions valid in this limit. 

As a first analysis, we search the minimal eigenfrequencies separation. Far from the crossing (i.e., for $|\delta\omega| \gg 2 g^2/|\Delta|$) the two eigenfrequencies are
\begin{align}
    \Omega_1&\,\simeq\,\omega_0-\delta\omega+\mathrm{Re}[G_{11}] \,\simeq\,\omega_0-\delta\omega+\frac{2g_1^2}{\Delta}\bigg( 1-\frac{\kappa^2/4-\omega_0^2}{\Delta^2}\bigg)\\
   \Omega_2&\,\simeq\,\omega_0+\delta\omega+\mathrm{Re}[G_{22}] \,\simeq\, \omega_0+\delta\omega+\frac{2\zeta^2 g_1^2}{\Delta}\bigg( 1-\frac{\kappa^2/4-\omega_0^2}{\Delta^2}\bigg)
\end{align}
\noindent which shows the cavity-induced frequency shift of the two modes. The crossing, defined by the minimal splitting between eigenfrequencies, occurs at
\begin{equation}
    \delta\omega=\frac{1}{2}\left(1-\zeta^2\right) \mathrm{Re}[G_{00}] = \frac{(1-\zeta^2)g_1^2}{\Delta} \bigg( 1-\frac{\kappa^2/4-\omega_0^2}{\Delta^2}\bigg)
\end{equation}
\noindent where the eigenvalues are given by
\begin{align}\label{eqs_eigen1}
    \lambda_\pm& 
    =\omega_0+\frac{1}{2}(\zeta\mp1)^2 \,G_{00}
    =\omega_0+\frac{(\zeta\mp1)^2g_1^2}{\Delta}\bigg(1-\frac{\kappa^2/4-\omega_0^2}{\Delta^2}\bigg)+\di \frac{(\zeta\mp1)^2g_1^2\kappa\omega_0}{\Delta^3}\\
    &=\Omega\mp \frac{2\zeta g_1^2}{\Delta}\bigg(1-\frac{\kappa^2/4-\omega_0^2}{\Delta^2}\bigg)+\di \frac{(\zeta\mp1)^2g_1^2\kappa\omega_0}{\Delta^3}
\end{align}
\noindent with a corresponding frequency splitting of $\delta_{12}=-\frac{4\zeta g_1^2}{\Delta}(1-\frac{\kappa^2/4-\omega_0^2}{\Delta^2})$.  The values reported in the main text correspond to a lower order expansion. 

As a second analysis, we consider $\delta\omega=0$, i.e., the bare frequencies crossing. The two eigenvalues are now given by
\begin{align}
    \lambda_+&=\omega_0\\
    \lambda_-&=
    \omega_0+(\zeta^2+1)\,G_{00}
    =\omega_0+\frac{2(\zeta^2+1)g_1^2}{\Delta}\bigg(1-\frac{\kappa^2/4-\omega_0^2}{\Delta^3}\bigg)-\di\frac{2(\zeta^2+1)g_1^2\kappa\omega_0}{\Delta^3}
\end{align}
\noindent which shows an unmodified upper branch and a downshifted and broadened  lower branch (for red detunings).

\subsection{Dark mode formation}
As in Sec.~\ref{sec_eigen}, we consider a simplified scenario with two coupled modes. The objective is to evaluate the cavity mode and show that, when the conditions for the dark mode formation are met, the upper branch decouples from it and would not appear in the cavity field spectrum. As before, we consider $g_2=\zeta g_1$ and, since we are interested in a region close to the avoided crossing, we approximate the couplings in Eq.~\ref{eq4s} by considering $\omega_{1}\simeq\omega_{2}\simeq\omega_0$ allowing us to rewrite Eq.~\ref{eq3s} as
\begin{equation}
    \mathbf{D}=\begin{pmatrix}
        \omega_\alpha&0\\
        0&\omega_\beta
    \end{pmatrix}+
    G_{00}\begin{pmatrix}
        1 & \zeta\\
        \zeta&\zeta^2
    \end{pmatrix} \, .
\end{equation}

\noindent Solving Eq.~\ref{eqs_dyn} in the Fourier domain one finds
\begin{equation}
    \mathbf{V}=(\di\omega\,\mathbf{I}+\di\mathbf{D})^{-1}\mathbf{V}_N=\mathbf{B}\mathbf{V}_N
\end{equation}
\noindent where $\mathbf{V}_N=(v_{1in},\, v_{2in})^{T}$, with $v_{1in}$ and $v_{2in}$ being the two modes heating rates. The matrix $\mathbf{B}$ is given by
\begin{equation}\label{eq_sol}
    \mathbf{B}=\frac{1}{\Delta_D}\begin{pmatrix}
        \di(\omega_2-\omega)+\zeta^2G_{00}&-\zeta G_{00}\\
        -\zeta G_{00}&\di(\omega_1-\omega)+G_{00}
    \end{pmatrix}=\frac{G_{00}}{\Delta_D}\begin{pmatrix}
        \zeta^2&-\zeta\\
        -\zeta&1
    \end{pmatrix}.
\end{equation}
Here, $\Delta_D=\mathrm{Det}[\di\omega\,\mathbf{I}+\di\mathbf{D}]$ and in the last step we have set $\omega=\omega_1=\omega_2$. 
Using Eqs.~\ref{eq_a} and \ref{eq_sol} we find
\begin{equation}
    a=\frac{\chi_c(\omega)\,g_1 G_{00}}{\Delta_D}\bigg[ \left( \zeta^2 v_{1in}-\zeta v_{2in}\right)+\zeta \left( -\zeta v_{1in}+v_{2in} \right)   \bigg]=0
\end{equation}
\noindent which shows that the dark mode always forms where the bare frequencies cross regardless of the value of $\zeta$.

\subsection{Coulomb coupling}\label{subscoulomb}
The Coulomb interaction provides a dynamical coupling between the particles degrees of freedom along with the steady state displacement of Eqs.~\ref{eqs_steady}-\ref{eqs_steady1}. This can be calculated by expanding to second order $H_C$ leading to an additional spring constant for each direction given by
\begin{align}
       k_z&=\frac{2 k_c}{r_{12}^3}\\
    k_{x,y}&=-\frac{k_c}{r_{12}^3}.
\end{align}
Focusing on the motion on the polarization planes, this contribution is included in the model by adding to the dynamical matrix $\mathbf{D}$ in Eq.~\ref{eqs_dyn} the following matrix
\begin{equation}
\mathbf{C}= g_c\begin{pmatrix}
\omega_{x1}^{-1} & 0 & (\omega_{x1}\omega_{x2})^{-1/2} & 0 \\
0 & \omega_{y1}^{-1} & 0 & (\omega_{y1}\omega_{y2})^{-1/2} \\
(\omega_{x1}\omega_{x2})^{-1/2} & 0 & \omega_{x2}^{-1} & 0 \\
0 & (\omega_{y1}\omega_{y2})^{-1/2} & 0 & \omega_{y2}^{-1}
\end{pmatrix}
\end{equation}

\noindent where $g_c=-k_c/2mr_{12}^3\equiv \frac{k_{x,y}}{2m}$. Two considerations can be made. First, it only contributes to the avoided crossing visible at $P_{\mathrm{tw2}}\simeq0.33$\,W in Fig.~3 of the main text, with a coupling that can be estimated to be $g_x\approx g_y\approx2\pi120$\,Hz thus representing only a small correction. Second, the cavity mediated interaction of Eq.~\ref{eq3s} has a periodic dependence on $\varphi$ with a period of $\pi$, however, the simultaneous presence of the Coulomb interaction breaks this periodicity which then becomes $2\pi$.

\section{Eigenfrequencies fitting}
As stated in the main text we model the experiment with nominal particle and cavity parameters. The latter are all measured independently while the former are consistent with previously reported measurements by our group~\cite{Ranfagni2021Two-dimensional}. This leaves four parameters to be estimated: the deviation from orthogonality between the cavity axis and the tweezers propagation direction ($\theta_z$), the position of the first particle ($1064$\,nm tweezer) in the cavity standing wave $\phi_1$, the fields ratio $\zeta$ and their relative phase $\varphi$.

For the evaluation of the angle $\theta_z$ we adopt a dedicated approach. Firstly, it is measured independently  by translating a single particle along the \emph{z}-axis through the cavity mode cross-section~\cite{Ranfagni2021Vectorial}.  The estimated value is $\mid\theta_z\mid=(1\pm0.5)\degree$. Notice that the method used does not allow to ascertain the sign of $\theta_z$, however, the final result does not depend on it, so we assume a positive sign. Despite the fact that $\theta_z$ is quite small, it is still sufficient to have a large impact since in the current experiment we cannot independently control the position of the two particles. For example, if we consider the first particle to be at a cavity node, i.e., $\phi_1=\pi/2$, for the second one we would have $\phi_2=\phi_1+k r_{12}\theta_z\simeq2.5$, corresponding to a shift of $160$\,nm along $z$. The particle is therefore more than halfway to the adjacent cavity antinode. This makes it clear that the direct measurement of $\theta_z$ is not sufficiently accurate, and its value needs to be varied within the experimental uncertainty.

The last point before discussing the fitting procedure concerns the reference value for $\Delta z$. This has been set by choosing the reference as the equilibrium position when the power on the second tweezer is $P_{\mathrm{tw2}}=280$\,mW, yielding $\omega_{z2}/2\pi\simeq23.3$\,kHz. Given the range of $P_{\mathrm{tw2}}$ exploited in the experiment, $\Delta z$ explores $\sim300$\,nm which correspond to an overall scan  of $\varphi$   by roughly   $110\degree$. Thus, this represents a non-negligible effect.

For the parameters fitting we proceed as follows. First, we initially fix $\theta_z$ and create three 2D maps $\mathcal{F}_i(\phi_1,\varphi)$  of the three splittings $\delta_{\alpha\beta}$, with $i=(x_1x_2,y_1y_2,x_1y_2)$, for each value of the fields ratio $\zeta$ in the range 0.28-0.45 with a resolution of 0.017. To generate the map, the eigenfrequencies of the dynamical matrix $\mathbf{D}$, including the Coulomb term, are calculated varying the power of the second tweezer $P_{\mathrm{tw2}}$. The three splittings are obtained as minimal differences between eigenfrequencies. We then use a least square method to extract the optimal values of $\phi_1$ and $\varphi$, that is we minimize
\begin{equation}\label{eqs_lsq}
    \mathcal{L}=\sum_i \bigg(\frac{\mathcal{F}_i-\delta_i}{\sigma_i}\bigg)^2=\sum_i\mathcal{L}_i
\end{equation}
\noindent where $\delta_i$ are the measured splitting values and $\sigma_i$ their standard deviation. Considering $\mathcal{L}$ as a function of $\zeta$ allows to select the optimal fields ratio. The procedure is then repeated for different values of $\theta_z$ within the experimental uncertainty.
\begin{figure}[ht]
\includegraphics[width=\textwidth]{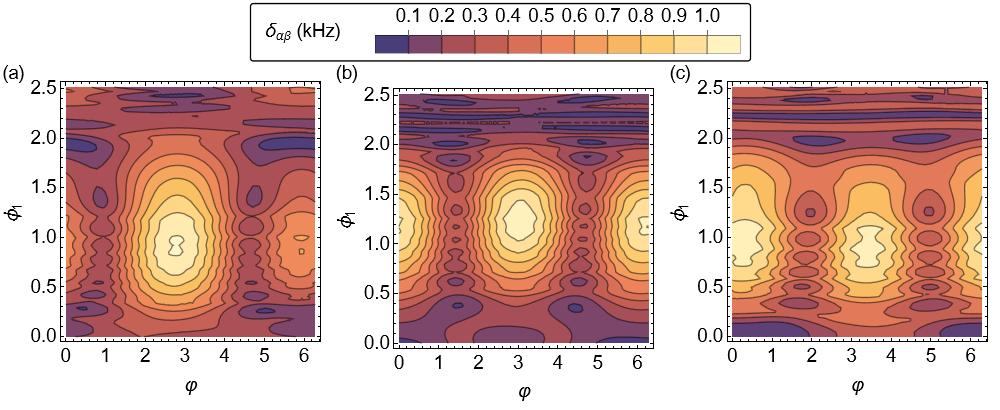}
\caption{Analytical 2D maps of the three splittings $\delta_{x1x2}$, $\delta_{y1y2}$ and $\delta_{x1y2}$ shown in a), b) and c) respectively as a function of the fields relative phase $\varphi$ and the position in the cavity standing wave $\phi_1$.
}\label{fig1sup}
\end{figure}
An example of the maps is shown in Fig.~\ref{fig1sup} for $\theta_z=1\degree$ and $\zeta=0.28$  which correspond to the initial estimate. It is immediate to see that $\delta_{y1x2}$ and $\delta_{x1y2}$, in Fig.~\ref{fig1sup}~b) and Fig.~\ref{fig1sup}~c) respectively,  approximately retain a periodicity of $\pi$ in $\varphi$, while $\delta_{x1x2}$ in Fig.~\ref{fig1sup}~a) does not. This is expected since $\delta_{x1x2}$ is the only one significantly affected by the Coulomb coupling 

At the end of the procedure the extracted parameters are $\theta_z=(0.75\pm0.13)\degree$,  a fields ratio $\zeta=0.35\pm0.015$ with a relative phase $\varphi=(172\pm3)\degree$ and a position along the standing wave of $\phi_1=1.225\pm0.015$.  This corresponds to the first particle being $\sim58$\,nm away from the node, and the second one $\sim 59$\,nm away on the opposite side.  

\begin{figure}[ht]
\includegraphics[width=\textwidth]{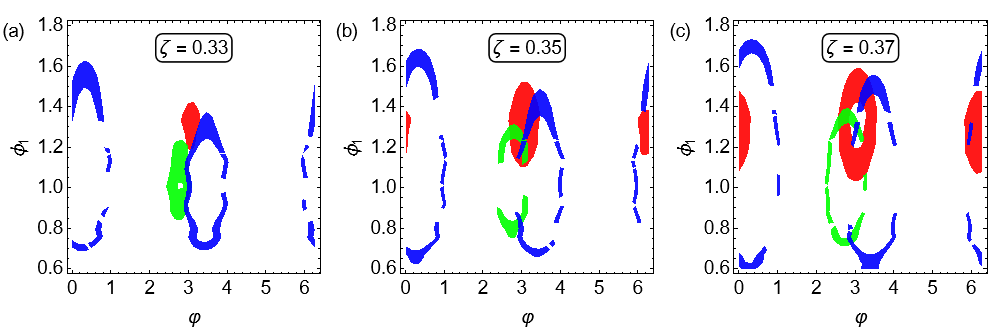}
\caption{Region plot for the individual contributions to $\mathcal{L}$ near the respective local minima. For all panels $\theta_z=0.75\degree$. The upper clipping is kept constant and equal to $1/\sigma_i$. The regions in red, blue and green correspond respectively to $\delta_{y1,x2}$, $\delta_{x1,y2}$ and $\delta_{x1,x2}$. The three panels correspond to different values of the fields ratio $\zeta$ with a) $\zeta=0.33$, b) $\zeta=0.35$,  c) $\zeta=0.37$. The second value yields the global minimum of $\mathcal{L}$, and indeed only in panel b) there is an intersection of all of the three plotted regions. 
}\label{fig2sup}
\end{figure}

To demonstrate the solidity of the procedure we show in Fig.~\ref{fig2sup}~b) the individual contributions $\mathcal{L}_i$, around the global minimum of $\mathcal{L}$, as a function of the map parameters $(\phi_1,\varphi)$. For each contribution we plot a region close to the local minimum $min(\mathbf{L}_i)$ with an upper clipping equal to $1/\sigma_i$. The global minimum occurs at the intersection of all three regions. In Fig.~\ref{fig2sup}~a) and Fig.~\ref{fig2sup}~c) we show the same plots with identical boundaries for the adjacent values of $\zeta$. It is clear that in neither case there is an intersection of all three regions.

\section{Analysis of additional contributions to the dynamics.}

\subsection{Overlap of the two trapping potentials}
Here we are interested in evaluating the perturbation on one particle due to the optical field used to trap the other. Since the two trapping positions are along the propagation direction, there will be both a gradient and a scattering force acting on a particle as a result of the off-focus optical field trapping the other. This modifies the total potential and influences the equilibrium position of the two particles. Importantly, to evaluate these contributions we do not require precise knowledge of the electric fields at the focus since the particles separation is much larger than the typical waist at the two foci. Thus, we can use the nominal Gaussian beam parameters that can be calculated for the aspheric lens doublet used to generate the optical tweezer. As stated in the main text, the two light fields are delivered through a polarization maintaining (PM) fiber which is first collimated, with a focal length $f=18.4$\,mm, and then rifocused by a second asphere, $f=3.1$\,mm. The fiber numerical aperture has been measured to be $NA=0.098\pm0.003$ which gives a waist of $w_{976}=0.53\pm0.02$ at $976$\,nm and $w_{1064}=0.58\pm0.02$ at $1064$\,nm. With these, the off-focus fields can be calculated with a standard Gaussian beam description.

The effect of the gradient force $F_{gr}$ can be calculated using the potential $H_{tw}=-\alpha \mid  \mathbf{E}(\mathbf{r})\mid^2/4$ while the radiation pressure force is given by $F_{rp}=P_{sc}/c$, where $P_{sc}$ is the total scattered power. Since we are interested in these forces in the linear part of the Gaussian beam expansion it is convenient to write approximate expressions. After some algebra one finds
\begin{align}\label{eqs_forces}
    F_{gr,j}&=\frac{2\alpha P_{tw,i}}{\pi c\epsilon_0}\frac{z_{R,i}^2}{w_i^2 z_{0,i}^3}\bigg( 1+\frac{3 z_i}{z_{0,i}} \bigg)=F_{gr0}+k_{gr} z_i\\
    \label{eqs_forces1}
    F_{rp,j}&=\frac{2P_{tw,i}\sigma_{sc,i}}{\pi c}\frac{z_{R,i}^2}{w_i^2z_{0,i}^2}\bigg(1+\frac{2z_i}{z_{0,i}}\bigg)=F_{rp0}+k_{rp}  z_i  
\end{align}
\noindent where the index refers to the two wavelengths, $z_{R,i}$ is the Rayleigh range, $\sigma_{sc,i}=\alpha^2k_i^4/(6\pi\epsilon_0^2)$ the Rayleigh cross section and $\alpha=\epsilon_0\chi V_s$ is the particle polarizability with $V_s$ the particle volume and $\chi$ its susceptibility. Given the experimental geometry, we have $z_{0,976}=r_{12}$ and $z_{0,1064}=-r_{12}$. These forces have to be compared with those resulting from the main optical potential. To do this, we use the experimental trap frequencies, thus we avoid the necessity of knowing the electric field at the foci. The steady state position is then given by $z_j=(F_{gr,i}+F_{rp,i})/m\omega_{z,j}^2$ and the modified trap frequencies are given by $\omega_{zf,j}=\sqrt{\omega_{z,j}^2-(k_{gr}+k_{rp})/m}$. When inserting in Eqs.~\ref{eqs_forces}-\ref{eqs_forces1} the parameters for the current experiment, we find that the steady state shift is larger for the particle in the $976$\,nm trap since the forces sum up. However, the estimated shift is of the order of $\sim100$\,nm, and it is just a few tens of nanometers for the particle in the $1064$\,nm trap. Thus, in both cases it remains negligible. As for the frequency shift, it is dominated by the softening gradient force contribution which for both particles remains smaller than $\sim200$\,Hz.

\subsection{Contribution to recoil heating}
Since the particles are illuminated by both beams there will be an additional recoil heating term. In general, the recoil heating rate for the center of mass motion is given by~\cite{Seberson2020Distribution}
\begin{equation}\label{eqs_recoil}
    \Gamma_{0}=\Lambda\frac{\omega_l}{\Omega}\frac{P_{sc}}{mc^2}
\end{equation}
\noindent where $c$ is the speed of light, $\omega_l=kc$, $P_{sc}$ is the total scattered power and the coefficient $\Lambda=(\frac{2}{10},\frac{1}{10},\frac{7}{10})$ takes into account the dipole pattern of the scattered light, $\Omega=(\omega_{x},\omega_{y},\omega_{z})$ are the different trap frequencies.  The scatter power can be written as $P_{sc}=\sigma_{sc}\epsilon_0c |E_0|^2/2$ where $E_0$ is the field at the trap center. This allows us to directly calculate the recoil $\Gamma_d$ at distance $d=r_{12}$, since we have $|E_{d}|=\zeta |E_0|$. In particular, the heating rate ratio is $\Gamma_{d}/\Gamma_0=\zeta^2=0.12$, assuming identical frequencies. Such a small ratio simply reflects the fact that $\Gamma$ depends on the intensity, while the coupling to the cavity depends on the field.

However, the total recoil per particle depends on the different wavelengths and, more importantly, on the orthogonal polarizations of the two trapping fields. For simplicity, we will assume identical parameters for the foci of the two traps, but the following can be generalized for different geometries. Thus, for the particle in the $1064$\,nm trap there  will be the usual recoil term plus the contribution due to the off focus $976$\,nm field
\begin{equation}\label{eqs_rqcoil1}
\begin{split}
    \Gamma_{1,0}&= \Lambda_1 \Xi_1 |E_1|^2/\Omega_1     \\
    \Gamma_{2,d}&=\Lambda_2 \Xi_2 \zeta^2 |E_2|^2/\Omega_1.  
\end{split}
\end{equation}
Here, the indexes $i=(1,2)$ identify the $1064$ and  $976$ traps respectively, the fields are considered at the center of the respective trap, $\Lambda_1=(\frac{2}{10},\frac{1}{10},\frac{7}{10})$, $\Lambda_2=(\frac{1}{10},\frac{2}{10},\frac{7}{10})$ and $\Xi_i=\epsilon_0 k_i \sigma_{sc,i}$. Similarly, for the particle in the $976$\,nm trap we have
\begin{equation}
    \begin{split}\label{eqs_recoil2}
    \Gamma_{2,0}&= \Lambda_2 \Xi_2 |E_2|^2/\Omega_2     \\
    \Gamma_{1,d}&=\Lambda_1 \Xi_1 \zeta^2 |E_1|^2/\Omega_2.  
\end{split}
\end{equation}
Thus, using Eqs.~\ref{eqs_rqcoil1}-\ref{eqs_recoil2} we can quantify the increase of recoil heating with the ratios
\begin{equation}
\begin{split}
    \frac{\Gamma_{2,d}}{\Gamma_{1,0}}&=\bigg(\frac{1}{2},2,1\bigg)\bigg(\frac{k_2}{k_1}\bigg)^5\zeta^2 (1+P_0 \Delta P) \\
     \frac{\Gamma_{1,d}}{\Gamma_{2,0}}&=\bigg(2,\frac{1}{2},1\bigg)\bigg(\frac{k_1}{k_2}\bigg)^5\zeta^2 (1+P_0 \Delta P)^{-1}
\end{split}
\end{equation}
\noindent where we set $P_{tw1}=P_0$ and $P_{tw2}=P_0+\Delta P$. Assuming identical powers, $\Delta P=0$,  we have $\Gamma_{2,d}/\Gamma_{1,0}=(0.04,0.16,0.08)$ and $\Gamma_{1,d}/\Gamma_{2,0}=(0.38,0.09,0.19)$ where the larger contribution is due to the polarization orientation.
    
\subsection{Optical binding}
Optical binding arises from the interference between the trapping fields and the scattered light. Thus, 
at each particle position there are two interfering terms, one for each wavelength. It is possible to show that, to the leading order, the coupling for the motion along \emph{z} vanishes and one is left with coupling terms $\propto x_1 x_2$ and $\propto y_1 y_2$ without mixed terms.

Along the $x$ and $y$ directions one can define an equivalent spring constant for each wavelength. For the particle in the $1064$\,nm trap we have
\begin{equation}\label{eqs_k1}
    k_{OB,i}=-\frac{\alpha^2k_i^3}{4\pi\epsilon_0 \,c\,d^2}\sqrt{I_{i,0}I_{i,d}}\,\,[1+(z_R/d)^2]^{-\frac{1}{2}},
\end{equation}
\noindent the index $i=(1,2)$ identify the $1064$\,nm and  $976$\,nm wavelenghts respectively.  On the other end, for the particle in the $976$\,nm trap we have
\begin{equation}\label{eqs_k2}
    k_{OB,i}=-\frac{\alpha^2k_i^3}{4\pi\epsilon_0 \,c\,d^2}\sqrt{I_{i,0}I_{i,d}}\,\sin\Phi_i
\end{equation}
\noindent where $\Phi_i=2k_i d-\arctan\frac{d}{z_R}$.  Eqs.~\ref{eqs_k1}-\ref{eqs_k2} are valid for both $x$ and $y$ directions and allow to calculate the coupling rates. Considering the parameters of the current experiment the estimated coupling is of the order of $g_{OB}/2\pi\sim1$\,Hz for all four terms  in Eqs.~\ref{eqs_k1}-\ref{eqs_k2} and is thus negligible. A detailed calculation for  Eqs.~\ref{eqs_k1}-\ref{eqs_k2} can be found in the supplemental document of Ref.~\cite{Deplano2024Coulomb}.

\section{Experimental setup}
\begin{figure}
\includegraphics[width=12.0cm]{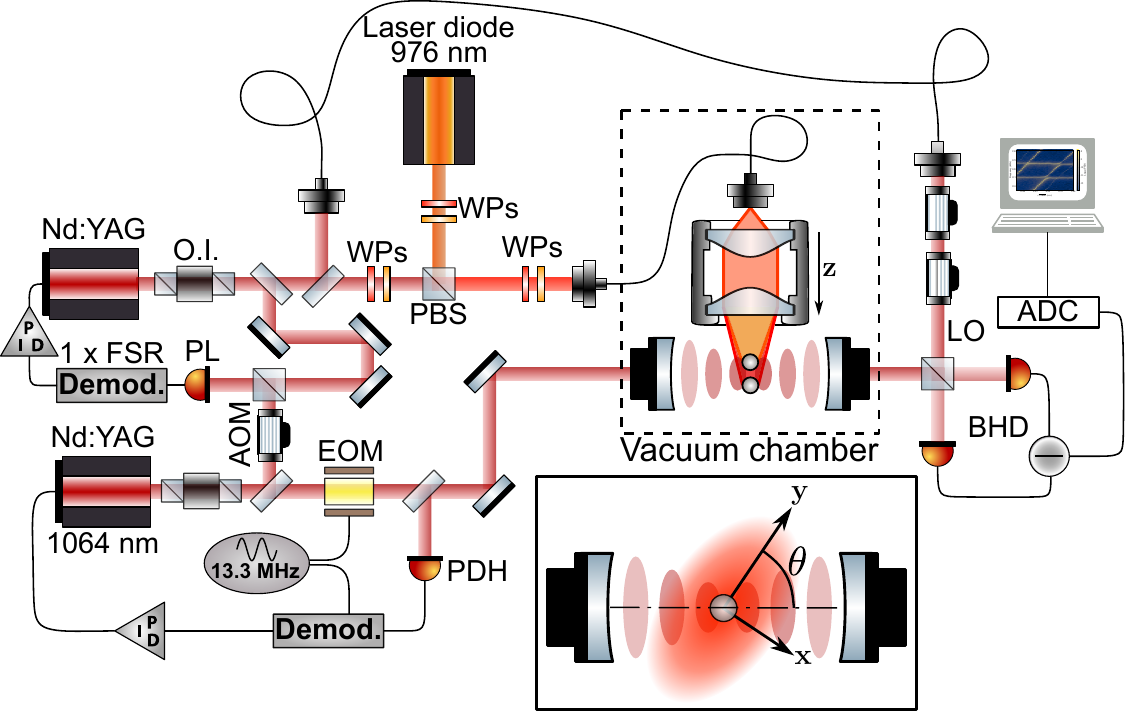}
\caption{Optical layout. O.I: optic isolator, WP: wave plate, PBS: polarizing beam-splitter, AOM: acousto-optic modulator, EOM: resonant electro-optic modulator, FSR: free-spectral range, PDH: Pound-Drever-Hall detection, LO: local oscillator, BDH: balanced heterodyne detection, ADC: analog-to-digital converter.  
}\label{fig3sup}
\end{figure}

The optical layout of the experiment is shown in Fig.~\ref{fig3sup}.  A first Nd:YAG at $1064$\,nm is used to lock a high finesse optical cavity by implementing a PDH lock scheme. A second Nd:YAG laser at $1064$\,nm and a laser diode at $976$\,nm are used to create a bichromatic tweezer creating two trapping sites separated by  $r_{12}=9\pm1\,\mu$m. The particles are positioned at the center of the cavity with a 3 axes translational stages. The two Nd:YAG lasers are phase locked with a frequency offset corresponding to one cavity free spectral range (FSR). The experimental setup is described in details elsewhere, additional information can be found in Refs.~\cite{Ranfagni2021Vectorial,Ranfagni2021Two-dimensional}. The loading method is described in Ref.~\cite{Calamai2021Transfer} while the bichromatic tweezer and the particles charge characterization can be found in Ref.~\cite{Deplano2024Coulomb}.

\bibliographystyle{nicebib}
\bibliography{main_bib}